# A MULTIVARIATE APPROACH TO SINGLE-MOLECULE THERMOPOWER AND ELECTRIC CONDUCTANCE MEASUREMENTS

Joseph M. Hamill, Christopher Weaver, and Tim Albrecht*

School of Chemistry, University of Birmingham, Edgbaston Campus, Birmingham B15 2TT, United Kingdom

*t.albrecht@bham.ac.uk

**ABSTRACT:** We report a method using scanning tunnelling microscope single molecular break junction to simultaneously measure and correlate the single-molecule thermopower and electrical conductance. In contrast to previously reported approaches, it does not require custom-built electronics and takes advantage of a trace-by-trace calibration of the thermal offset at the Au/Au contact, thus greatly facilitating thermoelectric measurements at the single-molecule level. We report measurements of three molecules: 1,4-di(4-(ethynyl(phenylthioacetate)) benzene, 1,8-octanedithiol, and 4,4'-bipyridine, and determine single-molecule Seebeck coefficients of 12(3), 5(2), and -5(2) µV K$^{-1}$, respectively. Furthermore, the method statistically correlates the Seebeck voltage offset, electrical conductance, and stretching displacement of the single-molecule junction, and allows for direct comparison with current-distance spectroscopy results obtained at constant bias.

**Introduction**

Thermoelectric energy generation is an environmentally friendly approach to improving the energy efficiency by converting waste heat into electricity. It has previously been estimated that approximately 63% of the global primary energy consumption is lost during combustion and heat transfer processes, with waste heat being a major contributor.[1,2] Efficient conversion of at least part of this loss into a usable form of energy would thus be an exciting prospect. While the theoretical efficiency of thermoelectric processes is not high and is bound by the Carnot limit (as well as other factors), they do have advantages, compared to other energy generation technologies, such as heat engines. For example, they are typically all-solid-state devices, which do not require cooling liquids or have any moving parts, and they are usually less bulky. This means they find application in areas, where maintenance is difficult or costly, such as in space travel or deep-sea exploration, or where power demands are low, for example in some sensing applications in remote environments.

Thermoelectric energy generation is rooted in the Seebeck effect,[3] where a temperature difference $\Delta T$ across a material induces a potential difference, the thermal voltage $\Delta V$. This is a consequence of the temperature dependence of the Fermi distribution, thus resulting in the redistribution of charge carriers in the material and in charge separation along the temperature gradient.[4] To generate a high $\Delta V$ for a given $\Delta T$ materials for thermoelectric applications must have a high Seebeck coefficient, defined by

$$S = -\frac{\Delta V}{\Delta T} \quad .^{[5]}$$

(1)

The efficiency of a thermoelectric material is related to its electric and thermal conductances, $G$ and $K$, via the figure of merit $ZT$ (assuming that the geometric factors for electric and thermal transport are the same):

$$ZT = \frac{GS^2T}{K} \quad .$$

(2)

Hence, it is clear from Eq. 2 that, to maximise $ZT$, $G$ and $S$ need to be as large as possible (thus maximising the power factor, $f = GS^2$), while $K$ must be minimized, for example by minimising phonon transport.[6]

Molecular materials equally lend themselves to such optimisation, due to their compositional and structural variety and synthetic flexibility. This is illustrated by eq. 3, which relates the thermopower to the transmission function $T(E)$ of a molecular junction[7].

$$S = -\frac{\pi^2 k_B^2 T}{3e} \frac{\partial ln(T(E))}{\partial E}\Big|_{E=E_F} \quad , \tag{3}$$

where $E$ is the energy, $E_F$ the Fermi energy, $k_B$ Boltzmann's constant, $T$ the temperature and $e$ the elemental charge.

Specifically, $S$ is determined by the gradient of the (logarithmic) transmission function $T(E)$, which can be altered by an appropriate choice of substrate material and anchor group, thus controlling the Fermi level alignment. In addition, it has also been shown that molecular design can affect the thermal conductance, for example via phonon interference, the introduction of heavy-atom side chains or the choice of anchor groups minimising phonon coupling to the substrates.[8] In combination, these features may offer the prospect of decoupling electric and thermal conductance in the spirit of the Wiedemann-Frantz law and potentially lead to further increases in $ZT$.

Experimental testing of theoretical predictions for molecular conductance and thermopower generally rely on two conceptual approaches, namely measurements at the single-molecule level, and in thin films, and one would expect these to be largely complementary. To this end, single-molecule measurements allow for more direct insight into structure-property relationships, and are unaffected by uncertainties around molecular surface coverage or structure, for example. On the other hand, they may also reveal a variance in the measured properties, which is not representative of thin-film devices, where high coverage and close packing in the molecular adlayer may reduce the structural variability. Thin-film measurements are more likely to represent structurally averaged characteristics as large number of molecules are sampled simultaneously.

However, measuring the thermoelectric properties of single molecules reliably is not trivial. Apart from challenges around addressing individual molecules immobilised on a surface, the thermal voltages induced by practical temperature gradients in molecular systems (perhaps 20-100 K) are small, e.g., in the region of 1-5 mV for $S = 50$ $\mu V$ $K^{-1}$. Scanning probe techniques, and in particular STM break junction (STM BJ) configurations, have been adapted to perform such measurements,[9,10,11,12,13] but this has typically involved customised hardware, purpose-built electronics, and sophisticated data acquisition schemes. For example, the pioneering work by Reddy et al. involved approaching a room temperature (RT) STM tip towards a heated substrate until a certain conductance value (0.1 $G_0$) was reached ($G_0$, the conductance quantum, is theoretically defined as $G_0 = 2e^2/h$, where $e$ is the elemental charge and $h$ is Planck's constant). The applied tip/substrate bias voltage and the current amplifier was then switched off, the tip retracted, and a voltage amplifier switched on to measure the thermal voltage directly during tip withdrawal.[14] A similar approach has been used by Malen and Segalman et al.[15] Widawsky et al. also used an STM BJ, first approaching the tip to a conductance value of 5 $G_0$, followed by tip retraction by a certain distance, holding the tip in position for 50 ms and another retraction. The first and last quartile of the 50 ms window were used to measure the molecular electric conductance at finite applied bias, while the bias was switched off during the middle 25 ms to record the thermal current. Finally, the measured thermal current and the electrical conductance were used to determine the thermal voltage.[16] Tao et al. performed thermoelectric measurements on DNA by soft-contact current/distance spectroscopy (the substrate was cooled), where molecular bridge formation during tip withdrawal was detected as a plateau in the tunnelling current (even though in their paper the authors do not seem to state how a plateau was identified during this process).[17] The withdrawal process was then stopped, a current/voltage ($I/V$) ramp performed between ±10 mV and tip withdrawal continued until the molecular junction broke. This measurement simultaneously provided the molecular conductance from the slope of the $I/V$ trace as well as the thermal voltage as the offset of the $I/V$ trace at zero (applied) bias. A conceptually simpler approach based on distance-dependent $I/V$ spectroscopy in a customised setup was employed by Yzambart, Agraït et al.[18] In their experiment, the (heated) tip was brought into contact with the substrate surface, then withdrawn in a stepwise manner (40-60 pm/step) and at each step an $I/V$ trace recorded (±10 mV, relative to a tip/substrate bias of 0.2 V). Again, both the conductance and the thermal voltage were extracted from the slope and the offset at zero bias, respectively.

In the present work, we employed a similar method, except that in our case the substrate was heated while the tip remained at RT, and we used the 'as supplied' logarithmic current pre-amplifier in the commercial STM setup, rather than custom-built electronics. Employing distance-dependent $I/V$ spectroscopy (STM BJ IV), we recorded multiple sets of $I/V$ traces to determine the voltage offset for $I$ = 0 nA, corresponding to either direct tip/substrate contact (close contact), a tip/molecule/substrate bridge, or direct tunnelling through a tip/substrate gap. The voltage offset at the Au/Au contact (~1 $G_0$) was then subtracted from all other voltage offsets in a given set, thereby reducing instrumental fluctuations and drift on a trace-by-trace basis. In combination with multivariate data analysis, this yielded an accurate measurement of the molecular thermal voltage and largely eliminated instabilities in the voltage offset that sporadically occurred over the duration of an experiment. This approach thus greatly facilitated the determination of single-molecule thermopower. It also provided distance-dependent data for the thermopower and electric conductance, as well as the break-off distance, and thereby enabled direct comparison with results from complementary constant-bias STM break-junction experiments, STM BJ.

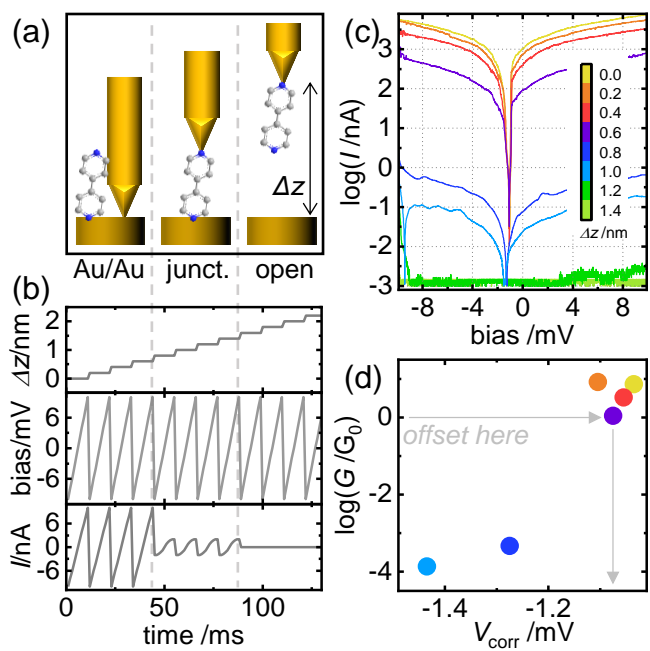

**FIG. 1**: Depiction of an STM BJ IV experiment and example data. (a) Au STM tip is retracted from the substrate while a bias is applied, and the current is measured. (b) In STM BJ IV mode the tip is retracted in steps, and at each step the bias is swept, yielding a series of $I/V$ traces across the junction. (c) Example series of $I/V$ traces from a single tip withdrawal at $\Delta T$ = 27 K, with four high-current traces across the Au/Au junction (yellow, orange, red, purple), two medium-current traces across molecule **3** (blue), and two traces at the noise level (greens). Note, asymmetry in these $I/V$ traces was a result of drift in the electronics and corrected in subsequent procedures. (d) Each trace in (c) can be parameterized by determining the displacement, $\Delta z$, within the series (colour scale), the differential conductance, $G$, and the voltage offset, $V_{corr}$. The trace closest to 1 $G_0$ (purple trace) defines the voltage and displacement offset corrections for that series (here $V_{corr}$ = -1.074 mV).

**Materials and Methods**

The three molecules used in this study were 1,4-di(4-(ethynyl(phenylthioacetate)) benzene (molecule **1**), 1,8-octanedithiol (molecule **2**), and 4,4'-bipyridine, (molecule **3**), and chosen to reflect different structure and anchor group chemistry as well as different thermopower magnitudes and signs. The samples were purchased from Sigma-Aldrich with purity grades of 97% (molecules **1** and **2**) and 98% (molecule **3**), respectively, and were used without further purification, cf. section S1.1 in the Supporting Information (SI) for further details.

Constant-bias STM BJ and distance-dependent $I/V$ spectroscopy measurements were conducted on an Agilent 5100 (Keysight) Scanning Tunnelling Microscope fitted with a logarithmic pre-amplifier and sample heating plate (Lake Shore Cryotronics Model 331, USA, controlling an Agilent Model N9647A, USA, sample plate). Substrate and STM tip were of Au; the tip was left at room temperature during all measurements. The main purpose of the constant-bias STM BJ experiments as a well-established technique was to determine $G_{mol}$ and $\Delta z_{mol}$ for each molecule, as described in section S1.2,[19, 20, 21,22] and to validate results obtained from STM BJ IV spectroscopy.

In STM BJ IV mode, $I/V$ sweeps were measured as a function of tip/substrate distance, as depicted in figure 1(a) and (b), in discrete steps of 0.2 or 0.3 nm (larger for molecule **1**, to ensure breaking of the junctions), from direct Au/Au contact ($\Delta z \leq 0$ nm) to large tip/substrate distances ($\Delta z >> \Delta z_{mol}$). At intermediate distances, when $\Delta z \approx \Delta z_{mol}$ or smaller, a molecule may be caught between the tip and the substrate until the distance becomes too large the junction eventually ruptures. In a typical experiment, $\approx$1000 withdrawals were performed in 25 steps at each $\Delta T$, yielding $\approx$25000 $I/V$ curves total.

At each step, an $I/V$ trace was recorded, typically between ±10 mV at a sweep rate of 2.0 V/s (see section S1.4 in the SI for detailed discussion on the effect of sweep rate, direction and voltage range), providing information on the thermal voltage (from the offset of $I$ at $V$ = 0 mV, *vide infra*), the electric conductance $G$ (from the slope of the $I/V$ trace, calculated from 41 data points centred around V = -5 mV), and the displacement of the tip relative to the substrate, $\Delta z$. Hence, for the three cases described above, one would expect $G \geq 1$ $G_0$ for sweeps across the Au/Au junction (yellow to purple traces in fig. 1(c)); $G \approx G_{mol}$ for sweeps across the Au/molecule/Au junction (blue traces); and $G \approx 0$ once the molecular bridge has been broken at sufficiently large $\Delta z$. By convention, $G$ was scaled by $G_0$ and plotted on a logarithmic scale.

The determination of the molecular thermopower, $\Delta V_{mol}$, required a more detailed analysis. For $\Delta T = 0$ and in the absence of random noise and drift in the electronics, $\epsilon_{STM}$, one would expect each $I/V$ trace to pass through $I = 0$ nA at $V = 0$ mV. For $\Delta T > 0$, the thermal voltage from the junction is superimposed on $V$, but in practice, $\epsilon_{STM}$ also contributes to $\Delta V$, cf. eq. (4). Detailed analysis has shown, however, that these interferences may be removed and the thermopower of the molecular junction, $\Delta V_{mol}$, be extracted, see section S1.3.2 for details. Briefly, the approach takes advantage of characteristic properties of the data, namely that a) random fluctuations may be averaged out using sufficiently large datasets, and b) electronic drift occurs on a different (slower) timescale than recording a set of $I/V$ traces, allowing for a voltage calibration on a trace-by-trace basis. Indeed, we found that the Au/Au contact with $G \approx G_0$ and $\Delta z = 0$ also serves as a suitable internal reference for the voltage offset of the $I/V$ traces and thus subtracted its value $V_{corr}$ from all $\Delta V$ values measured in $I/V$ traces in a given set of distances $\Delta z$, cf. fig. 1(d).

$$\Delta V = \epsilon_{STM} + \Delta V_{Au}(+\Delta V_{mol}) = \Delta V_{corr}(+\Delta V_{mol}) \qquad (4)$$

However, a number of points are worth noting. Firstly, the thermopower of the Au/Au contact itself has been found to be small, albeit somewhat dependent on the contact geometry.[23,24] To this end, Ludoph and van Ruitenbeek have reported a negative value of -0.5 μV/K for Au/Au contacts at $G_0$ and positive values up to +1 μV/K for $G$ up to 10 $G_0$. The former compares well with a value of approximately -1 μV/K reported by Agrait et al.[25], but is significantly smaller than the molecular thermopower values found for molecules **1** to **3** (see below). Secondly, while $\Delta V_{Au}$ determined at $G \approx G_0$ serves as an internal calibration here, it does not directly correspond to the thermal voltage of the contacts - namely the immediate atomic environment around the anchor groups - in the presence of a molecular bridge. Accordingly, we found that in the absence of a molecular bridge, after correcting $V_{corr}$, the average voltage offset was close to, but not exactly zero. We also explored whether differences in the load resistance would cause a systematic voltage offset, given that the load resistance during a STM BJ IV experiments changes by several orders of magnitude. To emulate this effect, we performed control experiments by bridging the tip and substrate electrodes with well-defined resistors in the range of $R_0 = 1/G_0$ and $R_{mol} = 1/G_{mol}$, and measured the voltage offset in a way similar to the experiments with molecular systems, cf. section S1.3.1. In the results, we did not observe a simple correlation between voltage offset and load resistance for the chosen range of values, fig. S5. However, there appeared to be small positive voltage offsets when comparing $R_{mol}$ against $R_0$. Crucially, while those factors may contribute to the observed absolute voltage offset, they seemed to be independent of $\Delta T$ within experimental error and are therefore unlikely to affect the determination of $S$, *vide infra*.

To proceed with the analysis, a three-dimensional ($\Delta z$, $G$, $\Delta V_{mol} = -\Delta V_{therm}$) scatter plot was produced and clustered using a Gaussian Mixture Model (GMM) with three clusters, cf. fig. 2(d) and section S1.3 in the SI.[13,20,26,27,28,29,30,31,32] This value was chosen because in most cases, the measurements were expected to include a cluster containing Au/Au junctions ($G > G_0$, yellow cluster in fig. 2 (c)), one cluster with molecular events (blue) as well as a separate cluster capturing noisy or poorly defined traces (red). Histograms of all $\Delta V$ values in the molecular, Au/Au, and noise clusters at given $\Delta T$ were then plotted, as shown in fig. 2 (d) for molecule **3** at $\Delta T$ = 27 K. In accordance with eq. (1), $S$ was then determined from a linear fit of all $\Delta V$ values plotted against $\Delta T$, as shown in fig. 3 (a) for molecule **3**, and figs. S7 (a) and (d) in the SI for molecules **1** and **2**. Replicates were performed on independent samples for each molecule and on separate days.

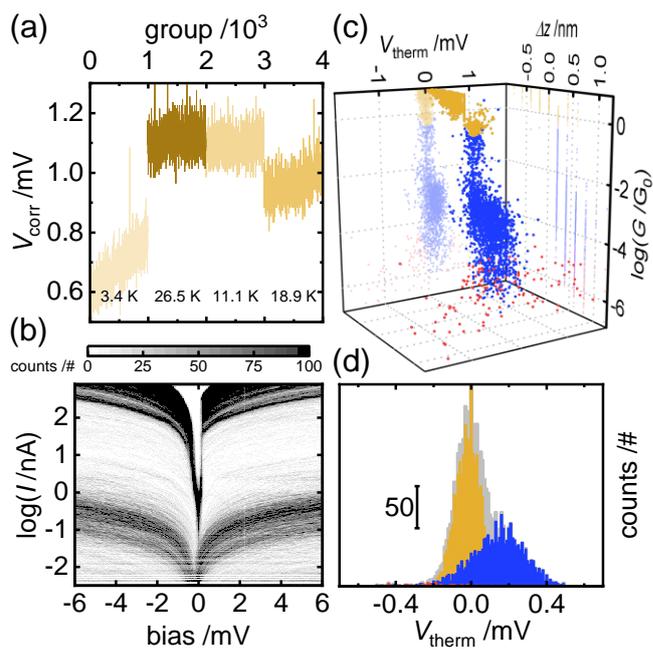

**FIG. 2**: Example STM BJ IV experiment performed on an adlayer of molecule **3**. (a) $V_{corr}$ for each withdraw group over the time of the experiment. (b) 2D current vs bias intensity plot after $V_{corr}$ is removed from each group. (c) 3D scatter plot of displacement, $\Delta z$, conductance, $G$, and voltage offset, $V_{therm}$, from each trace clustered using Gaussian mixture model into a Au/Au cluster (yellow), a molecular cluster (blue), and a noise cluster (red) (sweeps across open junctions had been removed). (d) 1D histograms of $V_{therm}$ for the three clusters above, and for the entire data set (grey) at $\Delta T$ = 27 K.

**Results and Discussion**

We now turn to the results obtained for molecules **1**-**3**, as well as "empty" Au/Au tunnelling junctions, following the above methodology. Fig. 3 (a-c) show results for **3**, similar data for **1** and **2** can be found in the SI, fig. S7. Panels (d) and (e) show the $\Delta V$ vs. $\Delta T$ plots, including the 95% confidence intervals, for all four configurations as well as the results obtained for $S$, including the individual replicates for each. The numerical results can be found in Table 1 (bottom rows), along with the power factor $f = G \cdot S^2$. The combined result for molecule **1** lies slightly below each of the individual replicates due to a constant offset between replicates 1 and 2, as seen in Fig. S7(a). Individually, both replicates have nearly identical results, except for a small constant offset. When the two results are combined, this offset reduces the magnitude of the slope overall.

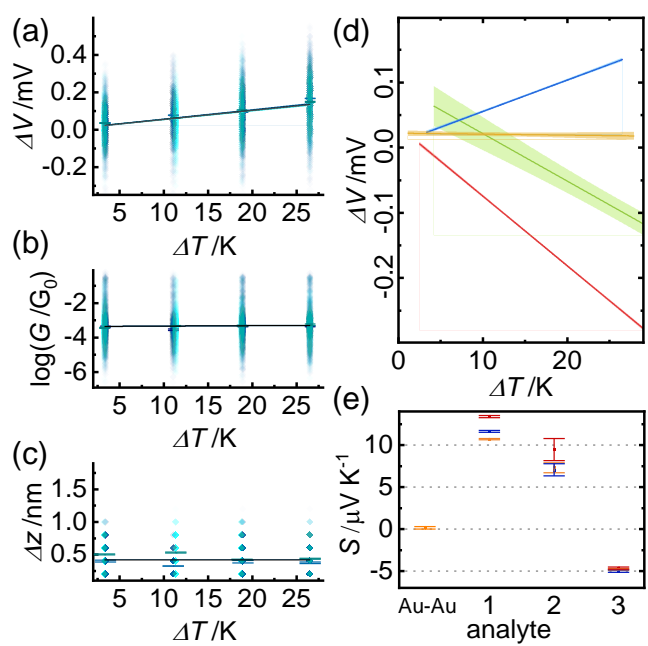

**FIG. 3.** (a) Scatter plots of $\Delta V$ vs $\Delta T$ measurements of molecule **3** with separate trend lines, and combined trend line with 95% confidence interval. (b) Scatter plots of $G$ vs $\Delta T$, and (c) $\Delta z$ vs $\Delta T$ for the same measurements, with $G_{mol}$ and $\Delta z_{mol}$ from each separate measurement calculated as the mean of a Gaussian fit of the distribution, and standard deviation as error bars. Black trend line is guide for the eye. (d) $\Delta V$ vs $\Delta T$ trend lines with 95% confidence intervals for molecules **1** (red), **2** (green), **3** (blue), and clean Au/Au (yellow). (e) Summary of $S_{mol}$ for all molecules in this study, and clean Au/Au junctions. Blue/red represent trial 1/2, and orange is the combined result. Error bars are standard error on the fit of the slope.

Focusing initially on panel (a), the $\Delta V$ vs. $\Delta T$ plot shows a positive correlation, in accordance with eq. (1). A linear fit yields a value $S$ = -4.82 ± 0.09 µV/K (in short 4.82(9) µV/K). $\log(G_{mol}/G_0)$ and $\Delta z_{mol}$ were determined to be -3.3(9) and 0.4(2) nm, as discussed in more detail below. Finally, there does not appear to be a strong or consistent correlation between $\log(G/G_0)$ and $\Delta T$ for the three molecules, as expected for tunnelling-based transport, and a weak negative one for $\Delta z_{mol}$ vs. $\Delta T$ with a slope of 1 pm/K or less for molecules **1** and **2**, but not for molecule **3** within experimental error, see Table S3 in the SI. A negative correlation could indicate that the stability of the junctions is reduced at larger $\Delta T$, perhaps because molecular bonds rupture more easily when the substrate is hotter or because of increased thermal drift. However, the observed effect is rather weak and a fuller explanation would clearly require further study.

Fig. 3 (d) also includes the $\Delta V$ vs. $\Delta T$ data for molecules **1** (red) and **2** (green) as well as data for the Au/Au tunnelling junction (yellow), namely in the absence of a molecular bridge ("empty" junction). $\Delta G_{mol}$ and $\Delta z_{mol}$ for the latter formally arise because the analysis method is agnostic as to whether a molecule is present in the junction or not. As a result, in an "empty" Au/Au junction, a small number of $I/V$ sweeps are still recorded in sub-$G_0$ conductance region, before the conductance drops below the cut-off defining value for $\Delta z_{mol}$ ($10^{-4.5}$ $G_0$, *vide supra*). Numerically, those feature a mean conductance of $\log(G_{mol}/G_0$= -3(1)) and a (short) mean displacement of 0.4 nm, consistent with pure tunnelling with a decay constant of ca. 10 nm$^{-1}$. After correcting for $V_{corr}$ the remaining voltage offset - effectively the offset shift between a single-point contact and a tunnelling junction at distance $\Delta z$ - is very small and remains unchanged as $\Delta T$ increases, with a slope of 0.2(2) µV K$^{-1}$.

For molecules **1** and **2**, the correlation between $\Delta V$ and $\Delta T$ is opposite in sign, compared to molecule **3**. The thermopower values have been determined as +10.69(8) µV/K and +7.3(6) µV/K, and

log($G/G_{mol}$) and $\Delta z_{mol}$ as -3(1) and 1.1(8) nm for molecule **1**, and -4(1) and 0.6(3) nm for molecule **2**, respectively.

Table 1: Results from current/distance spectroscopy at constant bias (top) and distance dependent *I/V* spectroscopy (bottom). Values for $\Delta z_{mol}$ and $G_{mol}$ are sample mean and standard deviation for all $\Delta T$ and all replicates; *S* values are slope and standard error of the slope for the trend line through all $\Delta T$ and all replicates. Note that the values for Au/Au have a different meaning, as discussed in the main text.

| *Molecule* | $\Delta z_{mol}$ | $G_{mol}$ | $S_{mol}$ | *f* |
|---|---|---|---|---|
| | nm | log($G\,\boldsymbol{G_0}^{-1}$) | μV K$^{-1}$ | aW K$^{-2}$ |
| **1** | 2.02(3) | -3.49(1) | | |
| | 1.1(8) | -3(1) | 10.69(8) | 6.7 |
| **2** | 1.0(5) | -4.3(5) | | |
| | 0.6(3) | -4(1) | 7.3(6) | 0.33 |
| **3** | 0.8(3) | -3.2(8) | | |
| | 0.4(2) | -3.3(9) | -4.82(9) | 0.86 |
| *Au/Au* | 0.4(2) | -3(1) | (-0.2(2)) | |

Before we return to the discussion of the (thermo)electric properties of the junctions, for validation we have also compared the STM BJ IV results with data obtained using the more widely established constant-bias STM BJ method. While the latter does not provide information on $\Delta V$, it does yield $G_{mol}$ and molecular break-off distances, in the present case with higher spatial resolution in our implementation of the STM BJ IV approach. The results are shown in fig. 4 (for molecule **3**) and figs. S2 and S3 (for molecules **1** and **2**) and summarised in Table 1 (top rows). In each case, the conductance/distance traces show the well-known feature around $G_0$, corresponding to direct contact between the tip and the substrate. As the distance is increased, the conductance either drops exponentially (for "empty" junctions) or enters a plateau-like feature (for molecular junctions). In the case of molecule **1** and molecule **3**, molecular junctions seem to dominate, suggesting that the hit rate is close to 100%; for **1**, the plateau feature is noticeably slanted. For molecule **2**, both exponential decays and plateau features are observed, and the hit rate is 40%, based on the area-type Gaussian fit of the break-off distance histogram. STM BJ data for molecule **3** appear to show two sub-populations with plateaus at higher and lower conductance, with $\log(G_{mol}/G_0)$ = -3.2(8) and approximately -4. The high-conductance population appears to be the dominant population. This is in-line with findings by Venkataraman et al., who rationalise the presence of two sub-populations based on different binding geometries and Fermi alignment in the junction.[35] Notably, we did not observe two clear sub-populations in the STM BJ IV data, see blue cluster in fig. 2 (c), but at present it is unclear whether this is due to a lack of resolution or differences in the pulling process.

For comparison, we have included mean $G_{mol}$ and $\Delta z_{mol}$ values and standard deviations from the STM BJ IV measurements in the 2D conductance-distance histogram in Fig. 4(a) (dark/light blue cross hairs: independent datasets 1 and 2; orange: combined values). They clearly match the conductance plateau of the high-conductance population, while the displacement appears to be somewhat shorter. Qualitatively similar behaviour is observed for molecules **1** and **2**, namely that the conductance values match within experimental error, while the mean displacements obtained from STM BJ IV are systematically shorter than those measured in STM BJ, Table 1 and section S1.2 in the SI for further details. The differences are approximately twice or thrice the size the piezo step between consecutive $I/V$ sweeps (either 0.2 nm or 0.3 nm, *vide supra*).

Hence, this observation can be rationalised, considering how the mean junction displacements are defined in the two cases: in STM BJ IV mode, the displacements are determined while the molecular junctions is still intact (before break-off, judged by the measured $G$ value). In STM BJ mode, on the other hand, the displacement histogram was calculated when the trace had reached the lower conductance cut-off, for example $10^{-4.5}$ $G_0$ (i.e., after the junction had ruptured). Thus, the observed systematic difference in $\Delta z_{mol}$ is unsurprising and the otherwise close correspondence between the two methods supports the   assignment of the molecular cluster in the STM BJ IV data that was ultimately used to derive $S_{mol}$.

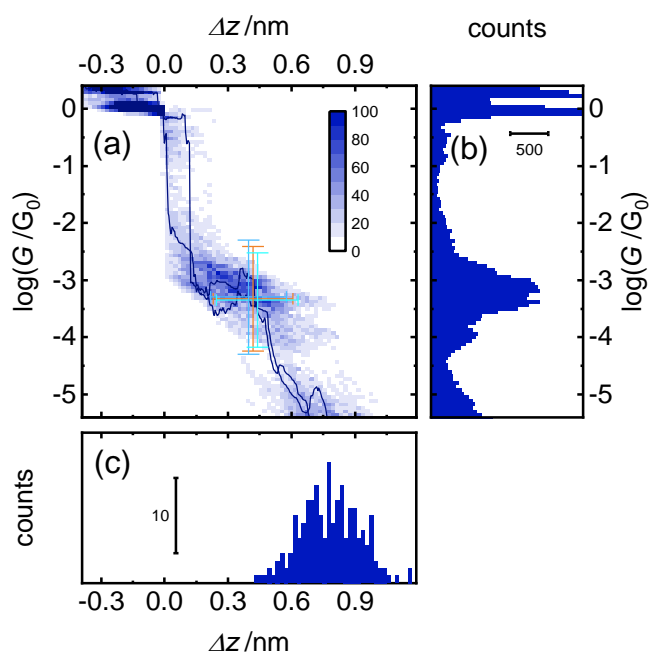

**FIG. 4**: Constant-bias STM BJ experiment performed on adlayer of molecule **3**. (a) 200 traces combined in a 2D conductance vs displacement intensity plot with two example traces (navy). Crosshairs are ($z_{mol}$, $G_{mol}$) from the STM BJ IV measurements with standard deviation (blues are trial 1 and 2, orange is combined). (b) 1D conductance histogram of all traces obtained in the constant-bias measurement. (c) 1D displacement histogram, determined at a conductance of $10^{-4.5}$ $G_0$.

Having consolidated the methodology, we will now briefly discuss our single-molecule conductance and thermopower results in a wider context. Generally, our conductance data are in good agreement with previous literature. For example, for **1**, we previously measured $\log(G_{mol}/G_0)$ = -3.74 using soft-contact current distance spectroscopy,[13] while Linke et al. found a conductance of $\log(G_{mol}/G_0)$ = -4.22 $\leq$ -3.92 $\leq$ -3.74 and a thermopower value 10.8 $\pm$ 9.5 μV/K,[33] which compares favourably to -3.49(1) and -3(1) and 10.69(8) μV/K, Table 1. In the case of molecule **2**, we found -4.3(5) and -4(1) for log($G_{mol}/G_0$) and $S$ = 7.3(6) μV/K, in agreement with log($G_{mol}/G_0$) = -4.3 in our previous work, again employing soft-contact current-distance spectroscopy (see reference 34 for a more detailed comparison). Moreover, Majumdar et al. determined the thermopower as 2.4 ± 0.4 μV/K,[15] which is somewhat smaller than the value of 7.3(6) we obtained in our experiments. Interestingly, for molecule **3**, we observed two conductance groups in STM BJ experiments, a more abundant one at log($G_{mol,H}/G_0$) = -3.2(8) and a less abundant one at ~-4, in line with previous work by Venkataraman et al.[16,35] They observed the high-conductance group at log($G_{mol,H}/G_0$) = -3.40 and the low-conductance group at log($G_{mol,L}/G_0$) = -4, with thermopower values of -8.4 μV/K and -7.5 μV/K, respectively. In our STM BJ IV measurements, we only observed one molecular cluster, for reasons discussed above, with a mean conductance of -3.3(9) and a thermopower of -4.82(9). This conductance value is in agreement with the high-conductance values above, even though our thermopower is formally lower than both values reported by Venkataraman et al.

**Conclusions**

Taken together, we find that the conductance, the sign and magnitude of the thermopower values for molecules **1-3** agree well with literature values, thereby validating previous results obtained on different measurement platforms. In comparison with the best performing molecular and solid-state materials, the thermopower values for those molecules are small,[36] likely due to the Fermi level being close to the middle of the HOMO/LUMO gap in these molecules. We note that in the case of molecular materials, however, the magnitude of $S$ can be significantly increased, for example by chemical design and quantum interference (QI),[37,38] which has been demonstrated experimentally and theoretically at the single-molecule and thin-film levels.[33] In the present work, we have mainly focused on the development and validation of STM BJ IV spectroscopy, as a new and comparatively facile way to measure single-molecule thermopower. We have also characterised the molecular conductance and break-off distances systematically in a range of temperature gradients, which we found to be negligible, and cross-validated STM BJ IV results with the more established constant-bias STM BJ method, finding close agreement in all three cases. On this basis, we are confident that our methodology can readily be applied to new molecular systems and facilitate the discovery of high-performing thermoelectric materials based on new paradigms in chemical design.

## References

[1] Cullen, J.M.; Allwood, J.M.  Theoretical efficiency limits for energy conversion devices. *Energy* **2010**, *35*, 2059-2069.

[2] Formann, C.; Muritala, I.K.; Pardemann, R.; Meyer, B.  Estimating the global waste heat potential. *Renewable and Sustainable Energy Reviews* **2016**, *57*, 1568-1579.

[3] Seebeck, T. J. Magnetische Polarisation der Metalle und Erze durch Temperatur-Differenz. *Abhandlungen der Königlichen Akademie der Wissenschaften zu Berlin*, Reimer, G., Ed., Deutsche Akademie der Wissenschaften zu Berlin: Berlin, **1822**, 265-373.

[4] Volta, A. XVII. On the electricity excited by the mere contact of conducting substances of different kinds. *Philos. Trans. Roy. Soc. London* **1800**, *90*, 403–431.

[5] Telkes, M. The Efficiency of Thermoelectric Generators. *J. Appl. Phys.* **1947**, *18*, 1116–1127.

[6] Lambert, C. J. Quantum Transport in Nanostructures and Molecules: An Introduction to Molecular Electronics. *IOP Publishing Ltd.*, 2021.

[7] Lambert, C.J.; Sadeghi, H.; Al-Galiby, Q. Quantum-Interference-Enhanced Thermoelectricity in Single Molecules and Molecular Films. *Comptes Rendus Physique* **2016**, *17*, 1084–1095.

[8] Famili, M.; Grace, I.; Sadeghi, H.; Lambert, C.J.  Suppression of Phonon Transport in Molecular Christmas Trees. *ChemPhysChem* **2017**, *18*, 1234–1241.

[9] Cui, L.; Miao, R.; Jiang, C.; Meyhofer, E.; Reddy, P. Perspective: Thermal and thermoelectric transport in molecular junctions. *J. Chem. Phys.* **2017**, *146*, 092201.

[10] Rincón-García, L.; Evangeli, C.; Rubio-Bollinger, G.; Agraït, N. Thermopower measurements in molecular junctions. *Chem. Soc. Rev.* **2016**, *45*, 4285–4306.

[11] Dubi, Y.; Di Ventra, M. Colloquium: Heat flow and thermoelectricity in atomic and molecular junctions. *Rev. Modern Phys.* **2011**, *83*, 131–155.

[12] Zhang, J.; Kuznetsov, A.M.; Medvedev, I.G.; Chi, Q.; Albrecht, T.; Jensen, P. S., Ulstrup, J. Single-Molecule Electron Transfer in Electrochemical Environments. *Chem. Rev.* **2008**, *108*, 2737–2791.

[13] Lemmer, M.; Inkpen, M.S.; Kornysheva, K.; Long, N.J.; Albrecht, T. Unsupervised vector-based classification of single-molecule charge transport data. *Nat. Commun.* **2016**, *7*, 12922.

[14] Reddy, P.; Jang, S.-Y.; Segalman, R.A.; Majumdar, A. Thermoelectricity in molecular junctions. *Science* **2007**, *315*, 1568–1571.

[15] Malen, J.A.; Doak, P.; Baheti, K.; Don Tilley, T.; Majumdar, A.; Segalman, R.A. Identifying the Length Dependence of Orbital Alignment and Contact Coupling in Molecular Heterojunctions. *Nano Lett.* **2009**, *9*, 1164–1169.

[16] Widawski, J.R.; Darancet, P.; Neaton, J.B.; Venkataraman, L. Simultaneous determination of conductance and thermopower of single molecule junctions. *Nano Lett.* **2012**, *12*, 354–358.

[17] Li, Y.; Xiang, L.; Palma, J.L.; Asai, Y.; Tao, N. Thermoelectric effect and its dependence on molecular length and sequence in single DNA molecules. *Nat. Comm.* **2016**, *7*, 11294.

[18] Yzambart, G.; Rincón-García, L.; Al-Jobory, A.A.; Ismael, A.K.; Rubio-Bollinger, G.; Lambert, C.J.; Agraït, N.; Bryce, M.R. Thermoelectric Properties of 2,7-Dipyridylfluorene Derivatives in Single-Molecule Junctions. *J. Phys. Chem. C* **2018**, *122*, 27198–27204.

[19] Gehring, P.; van der Star, M.; Evangeli, C.; Roy, J.J.L.; Bogani, L.; Kolosov, O.V.; van der Zant, H.S.J. Efficient heating of single-molecule junctions for thermoelectric studies at cryogenic temperatures. *Appl. Phys. Lett.* **2019**, *115*, 073103.

[20] Albrecht, T.; Slabaugh, G.; Alonso, E.; Al-Arif, M.R. Deep learning for single-molecule science. *Nanotechnology* **2017**, *28*, 423001.

[21] Huang, C.; Rudnev, A.V.; Hong, W.; Wandlowski, T. Break junction under electrochemical gating: testbed for single-molecule electronics. *Chem. Soc. Rev.* **2015**, 44, 889–901.

[22] Ratner, M.A. brief history of molecular electronics. Nature Nanotechnology **2013**, *8*, 378–381.

[23] Ludoph, B.; van Ruitenbeek, J.M. Thermopower of atomic-size metallic contacts. *Phys. Rev. B* **1999**, *59*, 12290.

[24] Tsutsui, M.; Morikawa, T.; Arima, A.; Taniguchi, M. Thermoelectricity in atom-sized junctions at room temperatures. *Sci. Rep.* **2013**, *3:3326*, 1–7.

[25] Evangeli, C.; Matt, M.; Rincón-García, L.; Pauly, F.; Nielaba, P.; Rubio-Bollinger, G.; Cuevas, J.C.; Agraït, N. Quantum Thermopower of Metallic Atomic-Size Contacts at Room Temperature. *Nano Lett.* **2015**, *15*, 1006–1011.

[26] Hamill, J. M.; Zhao, X.T.; Mészáros, G.; Bryce, M.R.; Arenz, M. Fast Data Sorting with Modified Principal Component Analysis to Distinguish Unique Single Molecular Break Junction Trajectories. *Phys. Rev. Lett.* **2018**, *120*, 016601.

[27] Cabosart, D.; Abbassi, M.E.; Stefani, D.; Frisenda, R.; Calame, M.; van der Zant, H.S.J.; Perrin, M.L. A reference-free clustering method for the analysis of molecular break-junction measurements. *Appl. Phys. Lett.* **2019**, *114*, 143102.

[28] Pearson, K. Contributions to the Mathematical Theory of Evolution. *Philosophical Transactions of the Royal Society A: Mathematical, Physical and Engineering Sciences*, The Royal Society, **1894**, *185*, 71–110.

[29] Jain, A. K.; Murty, M. N. & Flynn, P. J. Data clustering: a review. *ACM Computing Surveys* **1999**, *31*, 264–323.

[30] Mayor, M.; Weber, H.B. Statistical Analysis of Single-Molecule Junctions. *Angew. Chem.* **2004**, *43*, 2882–2884.

[31] Quan, R.; Pitler, C.S.; Ratner, M.A.; Reuter, M.G. Quantitative Interpretations of Break Junction Conductance Histograms in Molecular Electron Transport. *ACS Nano* **2015**, *9*, 7704–7713.

[32] Vladyka, A.; Albrecht, T. Unsupervised classification of single-molecule data with autoencoders and transfer learning. *Mach. Learn.: Sci. Technol.* **2020**, *1*, 035013.

[33] Miao, R.; Xu, H.; Skripnik, M.; Cui, L.; Wang, K.; Pedersen, K.G.L.; Leijnse, M.; Pauly, F.; Wärnmark, K.; Meyhofer, E.; Reddy, P.; Linke, H. Influence of Quantum Interference on the Thermoelectric Properties of Molecular Junctions. *Nano Lett.* **2018**, *18*, 5666–5672.

[34] Inkpen, M.S.; Lemmer, M.; Fitzpatrick, N.; Milan, D.C.; Nichols, R.J.; Long, N.J.; Albrecht, T. New Insights into Single-Molecule Junctions Using a Robust, Unsupervised Approach to Data Collection and Analysis. *J. Amer. Chem. Soc.* **2015**, *137*, 9971–9981.

[35] Kim, T.; Darancet, P.; Widawsky, J.R.; Kotiuga, M.; Quek, S.Y.; Neaton, J.B.; Venkataraman, L. Determination of Energy Level Alignment and Coupling Strength in 4,4'-Bipyridine Single-Molecule Junctions. *Nano Lett.* **2014**, *14*, 794–798.

[36] Roychowdhury, S.; Ghosh, T.; Arora, R.; Samanta, M.; Xie, L.; Singh, N. K.; Soni, A.; He, J.; Waghmare, U. V.; Biswas, K. Enhanced atomic ordering leads to high thermoelectric performance in $AgSbTe_2$. *Science* **2021**, *371*, 722–727.

[37] Bergfield, J. P.; Solis, M. A., Stafford, C. A. Giant Thermoelectric Effect from Transmission Supernodes. *ACS Nano* **2010**, *4*, 5314–5320.

[38] Sánchez, D.; López, R. Scattering Theory of Nonlinear Thermoelectric Transport. *Phys. Rev. Lett.* **2013**, *110*, 026804.

# A MULTIVARIATE APPROACH TO SINGLE-MOLECULE THERMOPOWER AND

# ELECTRIC CONDUCTANCE MEASUREMENTS

*Joseph M. Hamill, Christopher Weaver, and Tim Albrecht\**

*School of Chemistry, University of Birmingham, Edgbaston Campus, Birmingham B15 2TT, United*

*Kingdom*

*\*t.albrecht@bham.ac.uk*

**Supporting Information**

## 1   EXPERIMENTAL METHODS

### 1.1   SAMPLE PREPARATION

Oligo-phenylene ethynylene with thioacetate anchor groups (molecule **1**, Sigma-Aldrich 97%), 1,8'-octanedithiol (molecule **2**, Sigma-Aldrich 97%), and 4,4'-bipyridine (molecule **3**, Sigma-Aldrich 98%) were all used without further purification.

To adsorb the molecule on the substrate, first the Au substrate [MaTeck Au(111) single crystal disc, 11 mm diameter] was flame annealed for >60 s and allowed to cool under nitrogen. In the cases of molecules **1** and **3**, the substrate was then immersed in a 100 μM solution of the molecule in tetrahydrofuran (THF, Sigma-Aldrich 99.9% inhibitor free) for >30 min. For molecule **2**, the substrate was immersed in a solution of 1 ppb molecule **2** in THF for >30 min. In all cases, the substrate was removed from solution, rinsed in THF, and dried under nitrogen immediately before experiment assembly.

STM tips (Goodfellow, 0.25 mm diameter, 99.99% annealed) were electrochemically etched in a solution of equal parts hydrochloric acid (Merck Suprapur 30%) and ethanol (Fisher Scientific absolute 99.8%), with a potential of 14 V DC between the tip and a Au coil cathode.

## 1.2    Constant-bias single molecular break junctions

The scanning tunnelling microscope (STM) break junction (BJ) employs a STM to form junctions and measure the current of a single molecule bridging two gold electrodes.[1,2,3] Figs. S1(a) and (b) are cartoons of a single withdrawal during a single molecular break junction (SMBJ) experiment. The STM tip is brought into contact with the prepared substrate. When the STM tip is withdrawn at a constant rate (orange line) and constant bias (green line), a single molecule can bridge the tip and the substrate, forming a SMBJ, immediately before the junction breaks open. Fig. S1(b) depicts an idealized current trace (blue line) for a constant bias STM BJ. A typical withdrawal rate during a constant bias experiment is 8 nm s$^{-1}$; a typical bias is 100 mV. If the current between the tip and substrate is recorded during the withdrawal, the presence of the tip/molecule/substrate junction is apparent by a plateau-like feature in a plot of current vs displacement. If there is no molecule present in the junction, a plot of current vs displacement will exhibit only exponential tunnelling decay.

Panels (c)-(e) show experimental data from c. 200 traces of molecule **3**. Fig. S1(c) plots a 2D conductance-displacement histogram as an intensity plot of all traces overlaid. Two example traces are included for reference. The features in the region of 1 $G_0$ resulted from the atomic Au/Au point contact which is typically highly reproducible. The high point density region at $\approx 10^{-3.2}$ $G_0$ arose from the molecular features, which also gives rise to the peak feature in the 1D conductance histogram, fig. S1(d). Accordingly, a Gaussian fit yielded a mean molecular conductance, $G_{mol}$, of $10^{-3.2(8)}$ $G_0$.[4] The mean molecular plateau length was determined from a break-off distance histogram, fig. S1(e). The distance each trace reaches at a threshold conductance of, for example, $10^{-4.5}$ $G_0$, after the molecular plateau breaks, is recorded and histogrammed. A Gaussian fit of the distribution from the traces in Fig. S1(a) yielded a mean molecular plateau length, $\Delta z_{mol}$, of 0.8(3) nm. These constant-bias STM BJ results were used to validate the results from the STM BJ IV results presented in Fig. 4 for molecule **3** (main text) and Figs. Figs. S2 and S3 for molecules **1** and **2**.

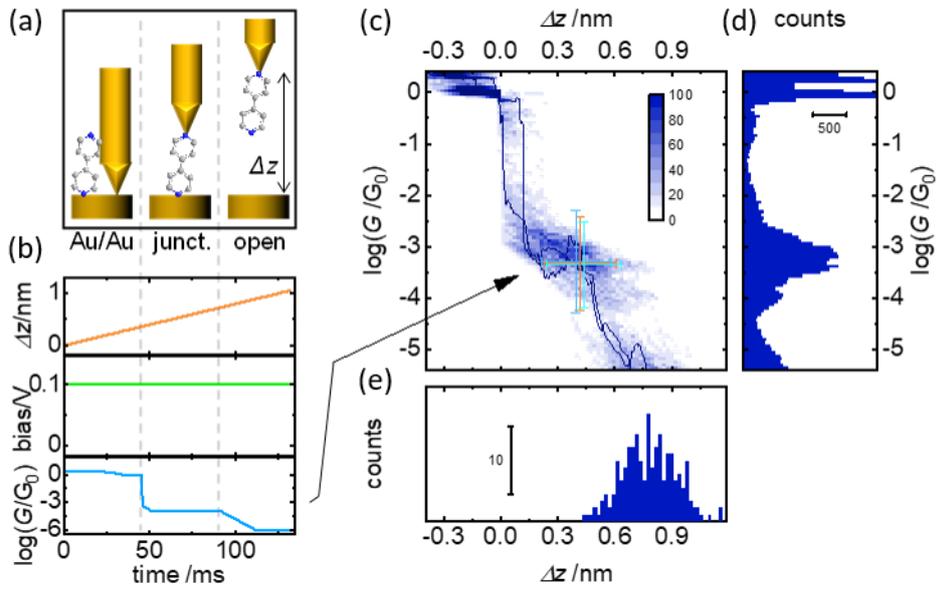

FIG S1. The STM BJ in constant bias mode. (a) The STM tip is brought into contact with the substrate and withdrawn repeatedly. (b) With a constant rate of withdrawal (orange line) and at constant bias (green line), a single current BJ trace is measured (blue line). (a) 2D conductance-displacement histogram with two overlaid example traces; (b) 1D conductance histogram with a peak feature at $10^{-3.2(8)}$ $G_0$; (c) break-off distance histogram with a peak at 0.8(3) nm.

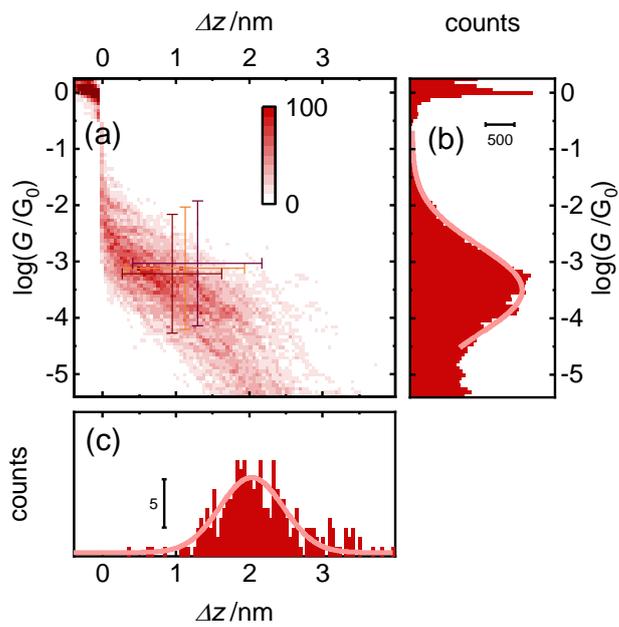

FIG S2. Constant bias results for molecule **1** (a) 2D conductance-displacement intensity plot of all traces. Cross hairs are trial 1 (dark red) and trial 2 (purple) and combined (orange) mean and standard deviation values of molecular cluster from STM BJ IV measurements. (b) 1D conductance histogram of all data in (a), with Gaussian peak fit (light red). (c) Break-off distance histogram, determined at conductance $10^{-5.1}$ $G_0$, with Gaussian fit (light red).

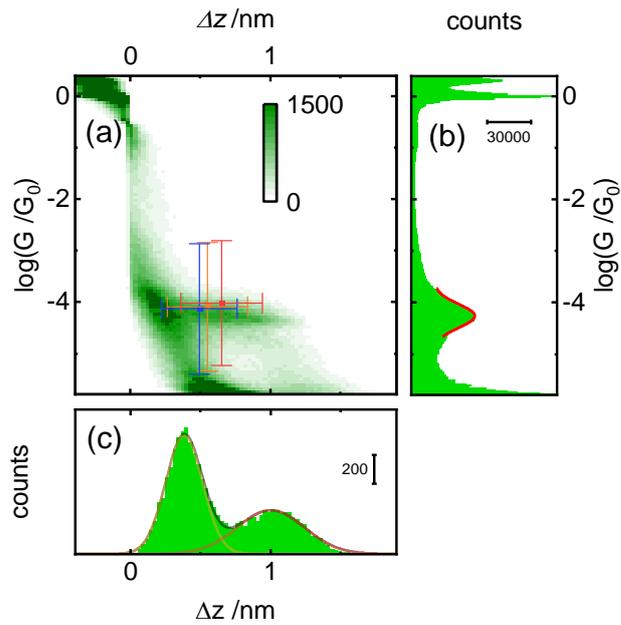

FIG S3. Constant bias results for molecule **2**. (a) 2D conductance-displacement intensity plot of all traces. Cross hairs are trial 1 (blue) and trial 2 (red) and combined (orange) mean and standard deviation values of molecular cluster from STM BJ IV measurements. (b) 1D conductance histogram of all data in (a), with Gaussian peak fit (red). (c) Break-off distance histogram, determined at conductance $10^{-5.0}$ $G_0$, with 2-peak fit to separate tunnelling (gold) and molecular (red) trace displacements. 2-peak fit was an area-type Gaussian fit, with centers at 0.4(3) and 1.0(5) nm, and relative areas of 60% and 40%, respectively.

## 1.3 Determining Seebeck voltage offset

A full-contact STM BJ IV approach was used in this study, whereby the STM tip was brought into full contact with the substrate at the beginning of each series of withdrawals, achieving a junction with conductance of >10 $G_0$. This form of STM BJ IV yields three main states within a single withdrawal: (1) the Au/Au junction, (2) the Au/molecule/Au junction, and (3) the open circuit. The Au/Au sweeps form the basis for the internal calibration of the applied bias which provides the resolution necessary for our method. After applying this internal calibration, the remaining $\Delta V$ was due to the Seebeck effect following the convention shown in Fig. S4.

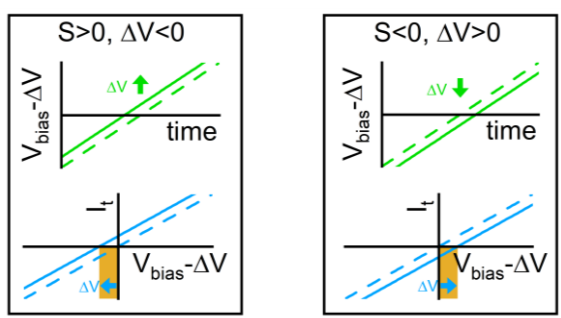

FIG S4. The influence of the Seebeck effect on $I/V$ sweeps. $\Delta V$ is negative (a) or positive (b) if the single molecular Seebeck coefficient is positive or negative, respectively. The superposition of the applied bias, $V_{bias}$ and $\Delta V$ results in a shift of the I/V equal to $\Delta V$.

### 1.3.1 3D Clustering

For each molecule, a set of ≈1000 withdrawal traces were acquired at various substrate temperatures. With 25 $I/V$ sweeps per trace, the resulting data set consisted of ≈ 25 000 $I/V$ sweeps. Of these sweeps, some were across the Au/Au junction, some were across the Au/molecule/Au junction, some were open circuit sweeps, and some contained poorly defined noise features. Sweeps across an open circuit were identified by their low standard deviation ($\sigma < 1$ pA) and eliminated from further analysis.

$V_{corr}$ was removed from all remaining traces following steps outlined in sec. 2.3 of the main text. Next, three parameters were calculated for each remaining trace, $\Delta V$, $G$, and $\Delta z$, as described in sec. 2.4. By reducing all remaining sweeps to three variables each sweep could be plotted on a 3D scatter plot like Fig. 2(c) of the main text. In this parameterized form, the sweeps can be clustered using an unsupervised machine learning clustering algorithm. For this study, a Gaussian mixture model (GMM) was selected, which is available as a built-in function in MATLAB. GMM suits the problem at hand because we expect each parameter in our scatter plot of parameterized $I/V$ sweeps to be normally distributed, or approximately so.[5,6,7,8] However, as shown in Fig. 2(c), this is only the case to a limited extent. While $G$ and $\Delta z$ distributions appeared skewed, the distribution of $\Delta V$ values was not. This would inevitably result in elongated clusters, for which algorithms such as kmeans, DBSCAN and GMM might perform poorly.[9] Thus, each variable was centred and normalised before applying the GMM, albeit with minimal effect on the results. We used this approach for all three molecules studied here, with a fixed number of three clusters in each analysis.

The clusters were then projected separately onto the $G$ and $\Delta z$ axes, histogrammed, and the histograms fitted to single Gaussian peaks. We then identified the class associated with the Au/molecule/Au $I/V$ sweeps for each temperature gradient, by choosing the cluster with mean values nearest to $G_{mol}$ and $\Delta z_{mol}$. In every case, the choice of cluster was unambiguous. Selecting for only molecular sweeps, all $\Delta V$s were used in the calculation of $S_{mol}$. Figs. S2(a) and S3(a), and Fig. 4(a) of the main text, show cross hairs where the mean and standard deviation of $G$ and $\Delta z$ from the molecular clusters from the STM BJ IV measurements map onto the constant bias measurements.

### 1.3.2 Electronic contribution to $V_{corr}$

To illustrate the variability in the 0 nA crossing voltage across the time span of an experimental data set, each $V_{corr}$, the offset voltage removed from each withdraw group due to electronic offsets and the Au contribution to the thermovoltage, was plotted as a time series in Fig. 2(a) of the main text.

As described in the main text, $V_{corr}$ was determined from the voltage offset of the last Au/Au $I/V$ sweep, identified by $G \approx 1$ $G_0$. The implicit approximation being applied in this correction was that the component of $V_{corr}$ which was due to $\epsilon_{STM}$ in Eq. 5 was independent of conductance. However, $G_{mol}$ was three or four orders of magnitude below 1 $G_0$ for all molecules in this study. To attempt to identify the component of $V_{corr}$ which was due to the resistance dependence of the logarithmic amplifier and identify any residual electronics contribution to $S_{mol}$, an average of 100 I/V sweeps across each of a series of resistors was plotted in Fig. S5, with equivalent conductances ranging from above 1 $G_0$ ($G_0 = 7.75 \times 10^{-5}$ S, 1 $G_0$ = 12.9 k$\Omega$) to below the sensitivity of our experiment at $10^{-9}$ $G_0$. For clarity, the offset due to the resistor nearest 1 $G_0$ was offset to 0 V. While there was no clear overall correlation between $V_{corr}$ and $G$, the values similar to the molecular conductances were positive relative to $V_{corr}$ at $G_0$. It is thus possible that such an effect contributed to the positive offset observed in Figs. 3(d). However, since this offset appeared to be constant, it had a negligible impact on $S_{mol}$.

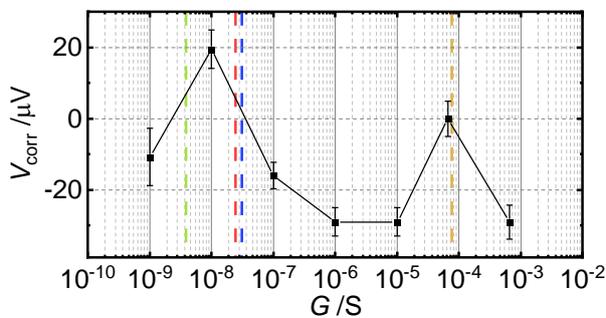

FIG S5. $V_{corr}$ dependence on conductance; $G_{mol}$ for molecules **1** (red), **2** (green), and **3** (blue), and 1 $G_0$ (orange) plotted as dashed lines; all offset so that 15 k$\Omega$ resistor (approx. 1 $G_0$) had offset of 0 µV; error bars are standard error of the mean with 100 samples each.

## 1.4 SENSITIVITY TESTING OF SWEEP RATE, SWEEP DIRECTION, AND BIAS WINDOW

To demonstrate that the Seebeck coefficient measured by our method was independent of the offset we observed in the Seebeck trend lines in Fig. 3(d), we conducted the same measurements with different sweep directions, sweep rates, and bias windows.

### 1.4.1 Seebeck coefficient dependence on sweep direction

Molecule **3** was measured sweeping the bias up from -10.0 mV to 10.0 mV to test the effect of sweep direction on the measured Seebeck coefficient. Both sweep directions yielded the same results within experimental error, as summarized in Table S1, and plotted in Fig. S6(a), and were in agreement with literature values.[10]

*Table S1. Summary of molecule **3** sweep direction dependence.*

|  | $\Delta z_{mol}$ | $G_{mol}$ | $S_{mol}$ |
|---|---|---|---|
| *Sweep direction* | nm | $\log(G\,G_0^{-1})$ | $\mu V\,K^{-1}$ |
| Up | 0.4(2) | -3.3(9) | -4.82(9) |
| Down | 0.4(5) | -3(1) | -7(1) |

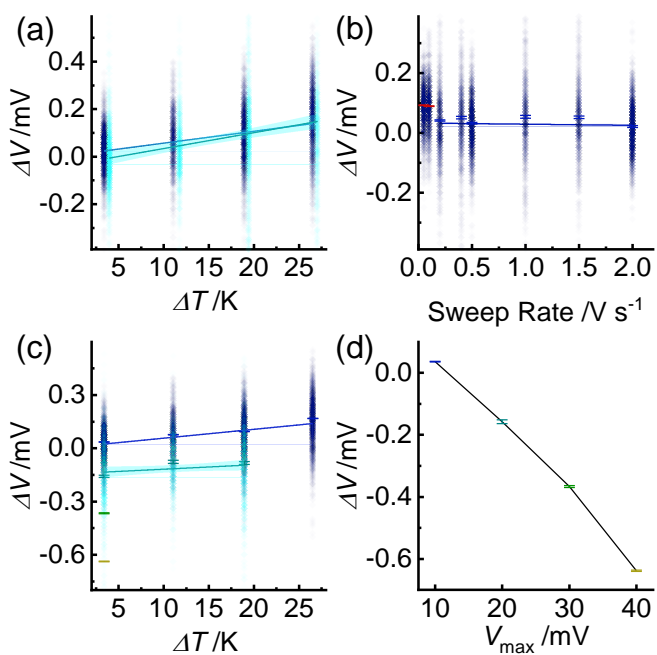

FIG S6. Sensitivity testing on molecule **3** results. (a) Sweep direction dependence (blue: sweep negative to positive; cyan: sweep positive to negative) with sweep window size (±10 mV) and sweep rate (0.2 V s⁻¹) (b) Sweep rate dependence at constant temperature gradient (3.4 K), scan direction (sweep negative to positive) and window size (±10 mV). (c) Sweep bias maximum dependence [10 mV (blue) and 20 mV (cyan)] at constant sweep direction (negative to positive) and sweep rate (2.0 V s⁻¹). Additional $\Delta V$ measurements with window maxima of 30 mV (green) and 40 mV (yellow) were measured at $\Delta T$ = 3.4 K. (d) $\Delta V$ vs scan bias maximum at $\Delta T$ = 3.4 K.

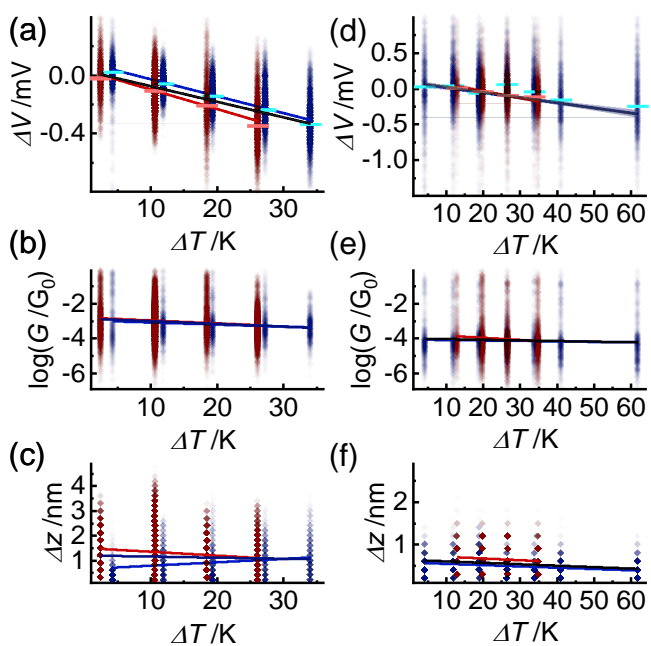

FIG S7. *ΔV*, *G*, and *Δz* vs *ΔT* plots for molecules **1** (a-c) and **2** (d-f). Slopes of black trend lines reported in Table S3. Red/blue points are from first/second measurements and red/blue lines are linear regressions of the first/replicated measurements. Light red and cyan cross hairs are means and standard deviations for individual temperature measurements.

### 1.4.2    Seebeck coefficient dependence on sweep rate

*ΔV* sweep rate dependence was tested on molecule **3** at constant *ΔT* of 3.4 K, and with a bias sweep of 10.0 mV to -10.0 mV, fig. S6(b). *ΔV* was constant for sweep rates above 0.2 V s$^{-1}$. Molecules **1**, **2**, and **3** were measured with a scan rate of 2.0 V s$^{-1}$.

### 1.4.3    Seebeck coefficient dependence on sweep bias maximum

In all experiments presented in the main text the bias was swept from 10.0 mV to -10.0 mV. To investigate whether there was any dependence of the bias range on the measured Seebeck coefficient, molecule **3** was measured also from 20.0 mV to -20.0 mV at three temperature gradients to obtain a Seebeck trend line, fig. S6(c) (dark blue). The trend line obtained for a bias range from 10 mV to -10 mV in light blue is shown for comparison. A single temperature gradient measurement was also measured for bias windows of 30.0 mV (dark green) to -30.0 mV and 40.0 mV to -40.0 mV (light green). Surprisingly, a plot of $\Delta V$ vs the maximum sweep bias, fig. S6(d), indeed an approximately quadratic dependence and at this point, we can only speculate as to where this effect is coming from; it may be related to an overshoot phenomenon in the applied bias and/or the time response of the preamplifier integration time. Crucially, however, we did not observe any detectable effect on the measured Seebeck coefficient.

*Table S2. Summary of the bias window dependence, for molecule **3**.*

| Bias max | $\Delta z_{mol}$ | $G_{mol}$ | $S_{mol}$ |
|---|---|---|---|
| mV | nm | log($G\,G_0^{-1}$) | μV K$^{-1}$ |
| 10 | 0.4(2) | -3.3(9) | -4.82(9) |
| 20 | 0.4(2) | -3(1) | -3(2) |

### 1.5 TEMPERATURE GRADIENT DEPENDENCE OF $\Delta z_{MOL}$ AND $G_{MOL}$

We further investigated whether there was any dependence of $\Delta z_{mol}$ and $G_{mol}$ on the applied temperature gradient, $\Delta T$, cf. Fig S7 and Table S3 for molecules **1** and **2**, and the main text for molecule **3**. In accordance with our expectations for electron or hole tunnelling as the dominant charge transport mechanism, there does not appear to be a strong, systematic dependence of $G_{mol}$ on $\Delta T$ for these molecules or indeed nominally for the Au/Au junction. The break-off distance $\Delta z_{mol}$ formally decreases, albeit very weakly, with increasing $\Delta T$ in all four cases.

*Table S3. Summary of temperature dependencies.*

| Molecule | Slope nm K$^{-1}$ | Std Err nm K$^{-1}$ | Slope log($G/G_0$) K$^{-1}$ | Std Err log($G/G_0$) K$^{-1}$ |
|---|---|---|---|---|
| **1** (OPE3) | -4.5E-3 | 3.7E-4 | -1.5E-2 | 4.9E-4 |
| **2** (1,8 ODT) | -3.3E-3 | 1.3E-4 | -3.7E-3 | 5.8E-4 |
| **3** (4,4 bpy) | -3.1E-5 | 1.6E-4 | 2.5E-3 | 7.9E-4 |
| *Au/Au* | -2.2E-3 | 1.0E-4 | 6.8E-3 | 6.4E-2 |

[1] Gehring, P.; van der Star, M.; Evangeli, C.; Roy, J. J. L.; Bogani, L.; Kolosov, O. V. & van der Zant, H. S. J. "Efficient heating of single-molecule junctions for thermoelectric studies at cryogenic temperatures," Applied Physics Letters, 2019, (115), 073103.

[2] Albrecht, T.; Slabaugh, G.; Alonso, E. & Al-Arif, M. R. "Deep learning for single-molecule science," Nanotechnology, 2017, (28), 423001.

[3] Ratner, M. "A brief history of molecular electronics," Nature Nanotechnology, 2013, (8), 378-381.

[4] Huang, C.; Rudnev, A. V.; Hong, W. & Wandlowski, T. "Break junction under electrochemical gating: testbed for single-molecule electronics," Chemical Society Reviews, 2015, (44), 889-901.

[5] Pearson, K. Contributions to the Mathematical Theory of Evolution *Philosophical Transactions of the Royal Society A: Mathematical, Physical and Engineering Sciences, The Royal Society,* 1894, (185), 71-110

[6] M. Mayor, H.B. Weber, Angew. Chem. 2004, 43, 2882-2884.

[7] Quan, R., Pitler, C. S., Ratner, M. A., Reuter, M. G. "Quantitative Interpretations of Break Junction Conductance Histograms in Molecular Electron Transport." ACS Nano 2015, 9, 7704–7713.

[8] Huang, C.; Rudnev, A. V.; Hong, W. & Wandlowski, T. "Break junction under electrochemical gating: testbed for single-molecule electronics," Chemical Society Reviews, 2015, (44), 889-901.

[9] Jain, A. K.; Murty, M. N. & Flynn, P. J., "Data clustering: a review," ACM Computing Surveys, 1999, (31), 264-323.

[10] Miao, R., Xu, H., Skripnik, M., Cui, L., Wang, K., Pedersen, K. G. L., Leijnse, M., Pauly, F., Wärnmark, K., Meyhofer, E., Reddy, P., Linke, H. "Influence of Quantum Interference on the Thermoelectric Properties of Molecular Junctions." Nano Lett. 2018, 18, 5666–5672.